\documentclass[applsci,article,accept,moreauthors,pdftex]{mdpi} 
\firstpage{1} 
\pubvolume{1}
\issuenum{1}
\articlenumber{0}
\pubyear{2023}
\copyrightyear{2020}
\datereceived{} 
\dateaccepted{} 
\datepublished{} 
\hreflink{https://doi.org/} 
\Title{Microscopic pair potentials and the physical properties of the condensed phases of parahydrogen}
\TitleCitation{Title}
\Author{Jieru Hu  and Massimo  Boninsegni\orcidA{}*}
\AuthorNames{Jieru Hu and Massimo Boninsegni}
\AuthorCitation{Hu, J.; Boninsegni, M.}
\address{\quad Department of Physics, University of Alberta, Edmonton, AB T6G 2E1, Canada}
\corres{Correspondence: m.boninsegni@ualberta.ca}
\abstract{Equilibrium physical properties of the solid and liquid phases of parahydrogen, computed by first principle computer simulations, are compared for different choices of pair-wise, spherically symmetric intermolecular potentials. The most recent {\em ab initio} potential  [Patkowski \textit{et al., J. Chem. Phys.}, 2008, \textbf{129}, 094304], which has a stiffer repulsive core than the commonly used Silvera-Goldman, yields results for structural quantities in better agreement with the most recent experimental measurements, while possibly overestimating the kinetic energy per molecule by as much as 10\%. Altogether, the comparison between theory and the available experimental evidence suggests that the  potential of Patkowski {\em et al.}  may be a better choice for simulations of condensed phases of parahydrogen at moderate pressure. }
\begin{document}
\section{Introduction}
\begin{figure}[h]
\centering
  \includegraphics[height=6cm]{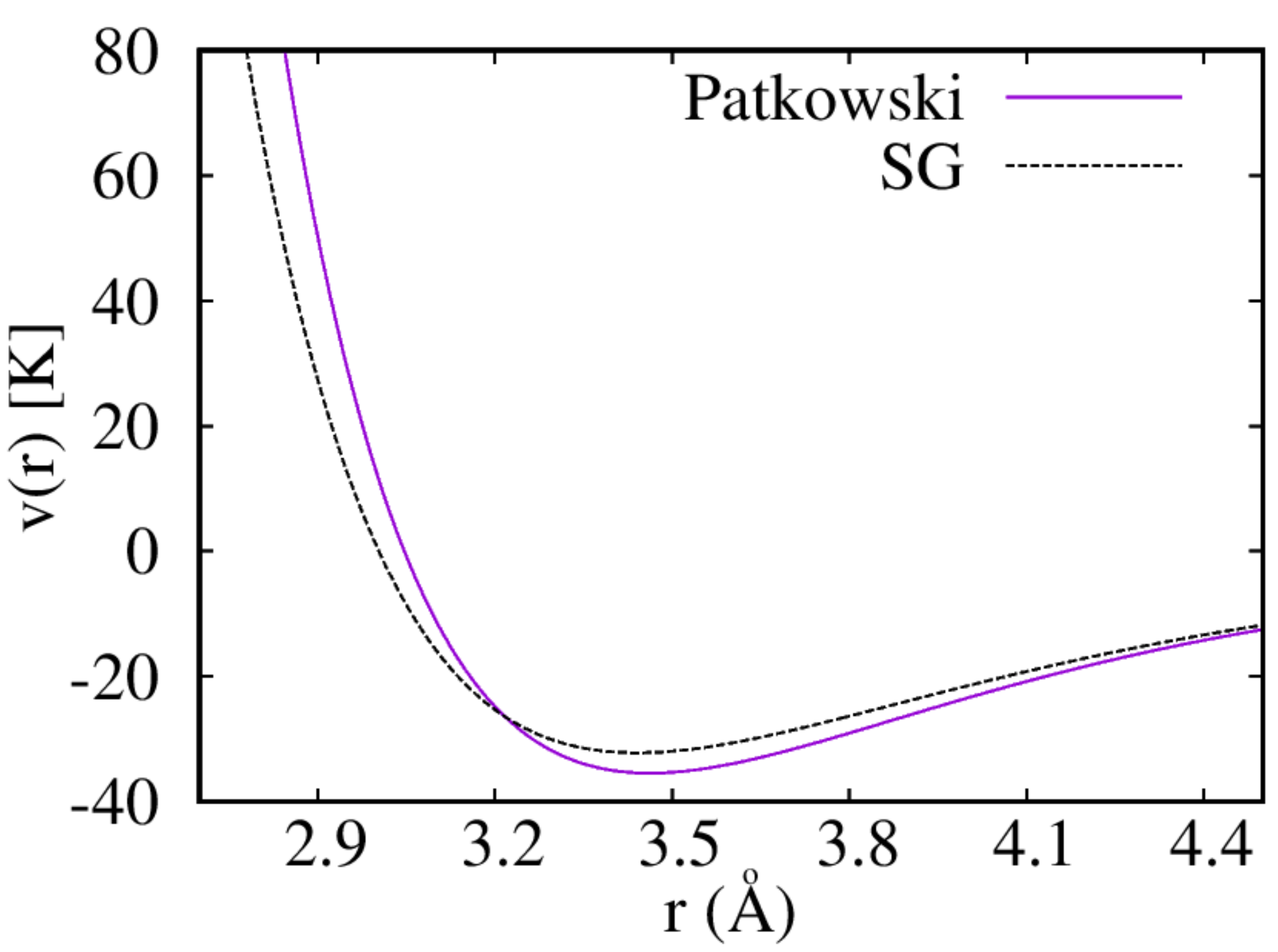}
  \caption{Comparison of Patkowski \cite{Patkowski2008} and Silvera-Goldman \cite{Silvera1978} (dashed line) intermolecular pair potential for {\em p}-H$_2$}
  \label{f0}
\end{figure}

Hydrogen, the simplest and most abundant element in the
universe, continues to motivate a lot of experimental and theoretical research,  because of its fundamental interest and technological relevance \cite{Silvera1980,Chabal1987,Myers1992,Mao1994,Nellis1995,Kohanoff2001,Guillot2005,Silvera2010,McMahon2012,Tozzini2013,Silvera2021}.
In particular, predicting quantitatively the equation of state (EOS) of the condensed 
phases of hydrogen, in a broad range of  thermodynamic conditions,  remains
a worthwhile theoretical goal, of  significant potential relevance to a wide
variety of pure and applied fields of science. 
\\ \indent
The {\em ab initio} theoretical study of the EOS of hydrogen in which  all the elementary constituents of the system (namely, electrons and protons), interacting through electrostatic Coulomb forces, are individually taken into account, constitutes a challenging quantum-mechanical many-body problem. Most calculations have been typically carried out within the framework of Density Functional Theory (DFT) \cite{Dharma-vardana1982,Barbee1989,Pickard2007} or quantum Monte Carlo (QMC) \cite{Ceperley1987,Natoli1993,Pierleoni1994,Magro1996}. Valuable as they are, {\em ab initio} approaches become impractical or unviable in specific physical regimes, due to the simultaneous presence of widely different energy scales; 
in those cases, one typically  resorts to approximations.
For example, when the system is in the plasma phase, at relatively high pressure and temperature, it is acceptable to treat the heavier protons as classical particles, and carry out semi-classical molecular dynamics or Monte carlo simulations, in which forces are evaluated either by DFT or QMC 
  \cite{Hohl1993,Kohanoff1997,Kitamura2000,Delaney2006,Attaccalite2008,Pierleoni2016,Mazzola2018}. \\ \indent 
On the other
hand, in molecular phases it is legitimate to adopt the
Born-Oppenheimer approximation and regard
individual hydrogen molecules as elementary particles, describing their interaction by means of a static potential, the
simplest choice being that of a pair-wise, central potential.
Besides being obviously unable to capture processes involving electronic transfer, such a crude model treats hydrogen molecules as spherical objects; this is known to be an oversimplification \cite{Diep2000}, even for the most nearly spherical (and abundant) isotope, namely parahydrogen ({\em p}-H$_2$). Equally absent are, of course, energy contributions of interactions involving, e.g., triplets
of molecules. 
\\ \indent
However, the use of static pair potentials renders the computation much faster and conceptually simpler than
{\em ab initio approaches}, allowing for the computation of thermodynamic properties of molecular hydrogen in
the condensed phase virtually 
without any uncontrolled approximation, e.g., by means of
QMC simulations. Thus, the development and testing of increasingly reliable pair potentials, affording a quantitatively accurate account for energetic and structural properties
of the condensed phase of molecular hydrogen in a wide range of thermodynamic conditions, is a worthwhile goal of quantum chemistry, one with likely impact in various areas of condensed matter physics and materials science.\\ \indent
The (mainly phenomenological) Silvera-Goldman (SG) \cite{Silvera1978} is
the most commonly adopted pair potential; it has been shown to afford
a reasonable, semi-quantitative description of low temperature structure and energetics
of the equilibrium liquid and solid phases of {\em p}-H$_2$ (the only species considered in this paper). Specifically, the computed (kinetic) energy per particle 
 \cite{Celli2000,Operetto2006,Gernoth2007,Boninsegni2009,Omiyinka2013} is in  agreement with {\em some} of the reported experimental measurements 
 \cite{Stewart1956,Schnepp1970,Driessen1979,Celli2000,Davidowski2006}, while the pressure is generally significantly underestimated. There are also quantitative differences between theory and experiment regarding structural correlations \cite{Celli2005}. 
Another popular pair potential is the Buck \cite{Norman1984}, featuring a slightly deeper
attractive well. These potentials are similar, in that the same general functional form is parametrized to reproduce some experimental data. These potentials yield an EOS in reasonable
agreement with experiment at moderate pressure (25 kPa); it is possible to ``tweak'' the SG potential, so that the theoretically predicted EOS will be in satisfactory agreement with experiment up to pressures of the order of $10^{8}$ kPa range \cite{Omiyinka2013,Moraldi2012}; these modifications are essentially {\em ad hoc}, however, and tailored to reproduce experimental values of only one or few physical quantities (usually the pressure).
\\ \indent
A different approach consists of determining the (time-independent) interaction between two {\em p}-H$_2$ molecules on the basis of first principle electronic structure calculations. The resulting potential energy surface reflects the lack of spherical symmetry of the molecule, so that the pairwise interaction depends not just on the distance between the molecules, but also on their relative orientation. An effective central potential is obtained by angular averaging.
The most recently proposed {\em ab initio} intermolecular potential for general use in the study of the condensed phases of {\em p}-H$_2$ is that of Patkowski {\em et al.} \cite{Patkowski2008}, henceforth also referred to as P. \\ \indent 
A theoretical study has established the accuracy of this potential in the determination of the viscosity and thermal conductivity of hydrogen in the gas phase \cite{Mehl2010}, but no systematic comparison  has yet been carried out of thermodynamic properties of the condensed phases of {\em p}-H$_2$, 
computed from first principles using the SG and P
potentials (shown for comparison in Fig. \ref{f0}). This is the goal of this work.

Specifically, we report results of large-scale computer simulations of the equilibrium  solid and liquid phases of  of {\em p}-H$_2$, based on QMC, making use of both the SG and Patkowski pair potentials. For completeness, we also report results of some calculations in which the Buck and the {\em ab initio} potential by Diep and Johnson \cite{Diep2000} were utilized. The purpose of this project is to determine whether recent {\em ab initio} potentials (and specifically the P) may represent a better choice than the SG, as a general-purpose potential to study condensed phases of {\em p}-H$_2$ in different settings, including droplets \cite{sindzingre1991,Mezzacapo2009,Boninsegni2019}, as well as in confinement \cite{Omyinka2014,Omyinka2016}.
\\ \indent
The results of this investigation paint a rather complex picture, in part due to inconsistencies between existing experimental data by different groups. In general we find that the {\em ab initio} potential, which have a stiffer repulsive core at short intermolecular distance, yield structural correlations, such as the static structure factor, with higher peaks, compared to those computed with the SG and Buck potential; for the liquid phase, this means better agreement with experiment, in fact close agreement with at least some of the reported experimental results. The pressure computed with the P potential is positive for both equilibrium phases (while the SG potential yields negative pressures),  with values that suggest that the incorporation of three-body terms into the P potential could bring the estimates yielded by it in agreement with experiment; this is because three-body terms, at the thermodynamic conditions of interest here, have been shown to result in a softening of the interaction, and therefore an overall reduction of the pressure \cite{Boninsegni1994}.
Finally, the P potential yields an approximately 10\% higher value of the kinetic energy per molecule than the SG; whether that means better or worse agreement with experiment really depends on what experiment one considers, i.e., there presently exist different, incompatible estimates. Therefore, it is necessary to resolve this outstanding experimental disagreement in order to make quantitative progress on the theoretical (computational) front. 
\\ \indent
This paper is organized as follows: in section \ref{model} we describe the model of interest and briefly review the computational technique utilized; in section \ref{res} we present our results, outlining our conclusions in section \ref{concl}.
\section{Model and Methodology}\label{model}
The system is described as an ensemble of $N$ point-like, identical particles with  mass $m$ equal to that of a {\em p}-H$_2$ molecule, and with spin zero, thus  obeying Bose statistics.  The system is enclosed in a cell of volume $\Omega$, so that $n=N/\Omega$ is the nominal density. The cell is shaped like a cube (cuboid) for simulations of the system in the liquid (solid) phase; periodic boundary conditions are used in all three directions. 
The quantum-mechanical many-body Hamiltonian reads as follows:
\begin{eqnarray}\label{u}
\hat H = - \lambda \sum_{i}\nabla^2_{i}+\sum_{i<j}v(r_{ij})
\end{eqnarray}
where the first (second) sum runs over all particles (pairs of particles), $\lambda\equiv\hbar^2/2m=12.031$ K\AA$^{2}$, $r_{ij}\equiv |{\bf r}_i-{\bf r}_j|$ and $v(r)$ is the pair potential which describes the interaction between two molecules, for which the various versions mentioned above have been utilized, for the purpose of comparing predictions for various thermodynamic quantities. 
\\ \indent 
We performed first principles QMC simulations of  the system described by Eq.  (\ref{u}), based on the canonical version \cite{Mezzacapo2006,Mezzacapo2007} of the continuous-space Worm Algorithm \cite{Boninsegni2006,Boninsegni2006b} based on Feynman's space-time approach to quantum statistical mechanics \cite{Feynman1965}  Since this methodology is extensively described in the literature, it will not be reviewed here. 
\\ \indent
We carried out the bulk of our simulations by treating {\em p}-H$_2$ molecules as {\em distinguishable}; this is because, although quantum exchanges have detectable effects on the momentum distribution of the liquid at equilibrium, near freezing \cite{Boninsegni2009}, they do not affect to a significant degree the quantities of interest here. Indeed, quantum exchanges are strongly suppressed in the bulk phases of molecular hydrogen, due to the relatively large radius of the repulsive core of the pair-wise interaction at short intermolecular separations \cite{Boninsegni2019b}. This is why all credible scenarios of possible {\em p}-H$_2$ superfluidity involve small cluster (comprising of the order of 30 molecules), not bulk phases \cite{Mezzacapo2009,Boninsegni2004,Boninsegni2013}.
\\ \indent
Details of the simulation are  standard; the short-time approximation to the imaginary-time propagator used here is accurate to fourth order in the time step $\tau$ (see, for instance, Ref. \cite{Boninsegni2005}). We have carried out numerical extrapolation of the estimates to the $\tau\to 0$ limit, and observed convergence of the thermal averages for a value of $\tau=2.5\times 10^{-3}$ K$^{-1}$ for all quantities of interest here, with the exception of  the pressure, which we computed using the well-known virial estimator \cite{Ceperley1995}; for this particular quantity, four times shorter a time step was required in order to arrive at a converged estimate. 
\\ \indent
Physical quantities of interest calculated in this study include, besides pressure and energetics, structural correlations such as the static structure factor. The kinetic energy per molecule $e_K$ and the static structure factor $S(q)$ can afford the most direct comparison of theory with experiment, as they are accessible through neutron scattering measurements \cite{Azuah1996,Celli2005,Davidowski2006}. We have carried out simulations for systems of varying size, the smallest (largest) comprising $N=256$ ($N=4096$) molecules. In general, we found that the numerical estimates of $e_K$ obtained on a simulated system comprising $N=256$ molecules are indistinguishable, within statistical uncertainties, from those obtained on systems of larger sizes, in both the liquid and the solid phase. On the other hand, some noticeable size dependence was observed for the other quantities; at any rate, the range of system sizes considered in this work gives us reasonable confidence in our ability to gauge finite-size effects, i.e., that the numerical values quoted herein are representative of the thermodynamic limit, within their statistical uncertainties.
\\ \indent
For the calculation of the potential energy and the pressure, we estimated the contribution of particles outside the main simulation cell by assuming a value of the pair correlation function $g(r)=1$, for $r$ greater than half the (shortest) cell side. Based on the observed quantitative consistency (within statistical uncertainties) of results for systems of different sizes, we contend that this procedure is numerically reliable.

\section{Results}\label{res}

We present here the results of our simulations for the liquid and solid phases of {\em p}-H$_2$, at few different conditions of temperature and density. We compare them with available experimental data; we include only reported experimental estimates for which values of density, temperature and pressure are furnished.
\subsection{Liquid}
Table \ref{t1} shows our numerical results for the kinetic energy per molecule and the pressure of the equilibrium liquid phase of {\em p}-H$_2$, obtained in simulations with different intermolecular potentials, at four distinct thermodynamic conditions for which experimental results are available. 
We first note that the results for the kinetic energy are similar for the phenomenological Silvera-Goldman and Buck potentials, {\em and nearly identical} for the first principle Patkowski and Diep-Johnson potentials, the values yielded by the first principle potentials being (rather consistently) $\sim$ 10\% higher. 
\\ \indent
Comparison with experiment for this particular physical quantity is complicated by the quantitative disagreement  between the two most recent experimental determination of the kinetic energy per molecule at the same (or at sufficiently close) thermodynamic conditions, namely that by Celli {\em et al.} \cite{Celli2000} and that by Davidowski {\em et al.}\cite{Davidowski2006} (shown in Table \ref{t1}); somewhat curious (and puzzling) is the fact that, while the result of Celli {\em et al.} is relatively close to the estimate yielded by the SG potential, that of Davidowski {\em et al.} agrees quantitatively with that based on the P potential. 
\begin{table}[t]
\small
  \caption{\ Simulation results for the kinetic energy ($e_K$, in K), and for the pressure (in bars), for the liquid phase of {\em p}-H$_2$ computed using different intermolecular pair potentials, at various conditions of density and temperature. Statistical errors (in parentheses) are on the last digit. Also shown are the most recent experimental estimates. The symbol $\sim$ is used if experimental uncertainties are not provided in the quoted reference. The experimental value marked with a $^\star$ is obtained by interpolating results of Table 1 in Ref. \cite{Celli2000} }
  \label{t1}
  \begin{tabular*}{0.48\textwidth}{@{\extracolsep{\fill}}lll}
    \hline
    Potential  & $e_K$ (K) & Pressure (bars) \\
     \hline
    \multicolumn{3}{c}{$n=0.02252$ \AA$^{-3}$, $T=15.7$ K}\\
    Silvera-Goldman &61.7(2) &$-15.3(5)$ \\
    Patkowski {\em et al.} & 66.7(2) & $7.4(5)$\\
        Experiment\cite{Celli2000} &$60.0(6)$ &$0.24(1)$ \\
    \hline
    \multicolumn{3}{c}{$n=0.02235$ \AA$^{-3}$, $T=16.5$ K}\\
    Silvera-Goldman & 61.8(2) & $-14.2(3)$\\
    Buck &62.6(1) &$-27.1(5)$ \\
    Patkowski {\em et al.} &67.1(2)  &6.9(3)  \\
    Diep-Johnson & 67.0(1) & 20(1) \\
    Experiment\cite{Davidowski2006} &$\sim67.8$ &$\sim 1$ \\
    Experiment\cite{Celli2000} &$60.3(6)$ &$0.33(2)^\star$ \\
        \hline
        \multicolumn{3}{c}{$n=0.02295$ \AA$^{-3}$, $T=17.1$ K}\\
    Silvera-Goldman &64.1(1)  &9.6(4)  \\
    Patkowski {\em et al.} &69.7(2)  &39.1(5) \\
    Experiment\cite{Celli2001} & $-$ &$29.9(1)$ \\
    \hline
    \multicolumn{3}{c}{$n=0.02204$ \AA$^{-3}$, $T=19.33$ K}\\
    Silvera-Goldman & 63.0(1) & $0.2(3)$\\
    Patkowski {\em et al.} & $68.3(2)$ & 22.1(5)\\
    Experiment\cite{Celli2001} &$63(3)$ &$17.4(5)$ \\
    \hline
  \end{tabular*}
\end{table}
To our knowledge, this outstanding, significant inconsistency between different experimental results is yet to be resolved. In general, it appears as if the SG potential may yield results for $e_K$ is closer agreement with experiment, although in some cases the size of experimental uncertainties prevents one from making a definitive conclusion. Thus, the situation at the present time is unclear; it is worth pointing out that the extraction of $e_K$ from the measured momentum distribution is a rather delicate operation, and that similar disagreements between theory and experiments and/or different experimental estimates have been reported in other contexts, e.g., in helium mixtures \cite{Diallo2006,Boninsegni2018}.
\\ \indent
The pressure is consistently underestimated by the SG (and the Buck) potential, and overestimated by the P (and Diep-Johnson). The difference between the results yielded by the two potentials can be attributed in part to {\em a}) the softer repulsive core at short distance, and {\em b}) the presence in these potentials of an attractive term, proportional to the inverse ninth power of the intermolecular separation, which accounts in an effective way for three-body interactions. This term is not present in the P and Diep-Johnson potentials; this is because, in an {\em ab initio} approach, non-additive terms describing the interaction of groups of three or more molecules are generally included as separate contributions to the total potential energy. We come back to this point in Section \ref{concl}, when we discuss more extensively the possible role of three-body terms; for the moment, we note that, among the various pair-wise intermolecular potentials considered in this work, the Patkowski potential furnishes estimates of the pressure that are closest to the experimental results, as shown in Table \ref{t1}. 
\begin{figure}[h]
\centering
  \includegraphics[height=6.8cm]{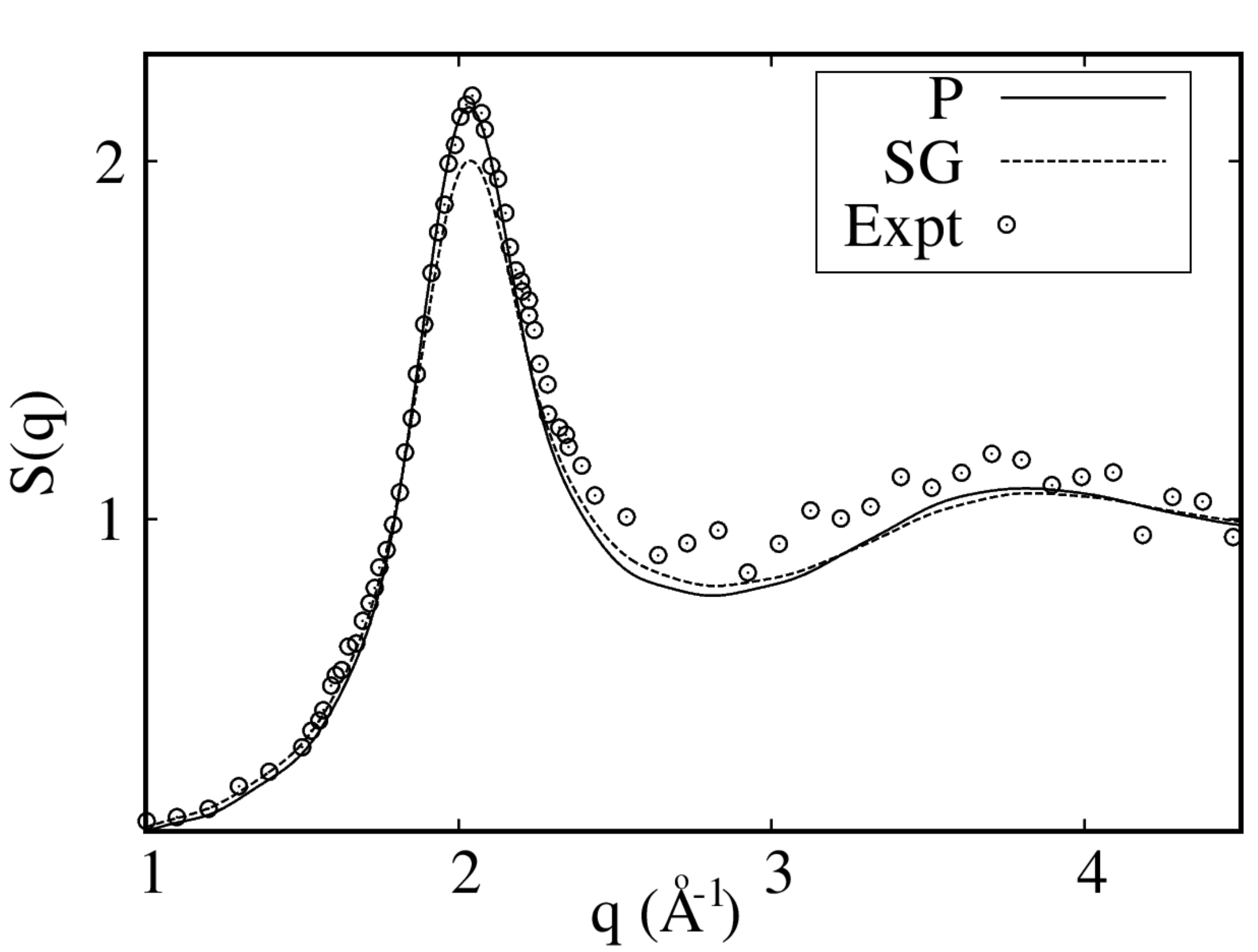}
  \caption{Static structure factor of liquid {\em p}-H$_2$ at density $n=0.02295$ \AA$^{-3}$ and temperature $T=17.1$ K, computed in this work using the SG (doted line) and P (solid line) intermolecular pair potentials. Also shown (circles) are experimental results from 
 Ref.\cite{Celli2005}. Experimental uncertainties are not shown for clarity but they are typically of the order of 0.1 for $q \ge 3 $ \AA$^{-1}$ }
  \label{f1}
\end{figure}
\\ 
More cogent information is provided by the comparison of structural correlations, calculated with the two different potentials, for which a direct comparison with experiment is possible. Fig. \ref{f1} shows the static structure factor $S(q)$, obtained from the numerically computed pair correlation function $g(r)$ through
\begin{equation}
    S(q) = n \ \int dr\ r\ g(r) \ sin(qr)
\end{equation}
which takes advantage of the isotropic character of the fluid phase. The agreement between experimental and simulation results based on the P potential is excellent, at least for $q\le 2.5$ \AA$^{-1}$, while the SG potential yields a significantly lower peak, as already noticed in Ref. \cite{Celli2005}. On the other hand, the SG potential and the P potential yield very similar results for $q\ge 2.5$ \AA$^{-1}$, and there may be a significant difference with experimental results, although that is not clear, given the uncertainties quoted on the experimental data (not shown for clarity in Fig. \ref{f1}).
\\ \indent
Altogether, the P potential seems to afford superior agreement with experiment than the SG potential, at least for the static structure factor, pointing to a more accurate description of the local environment experienced by {\em p}-H$_2$ molecules, at least in their immediate vicinity. Indeed, the agreement with experiment afforded by the P potential appears rather satisfactory, comparable, for example, to that yielded for the superfluid phase of $^4$He by the commonly adopted Aziz pair potential \cite{Ceperley1995}.
\subsection{Solid}
\begin{table}[t]
\small
  \caption{\ Simulation results for the kinetic ($e_K$) and total ($E$) energy per molecule (in K), and for the pressure (in bars) in the crystal (hcp) phase of {\em p}-H$_2$ at temperature $T=4$ K and density $n=0.0261$ \AA$^{-3}$, computed using different intermolecular pair potentials. Statistical errors (in parentheses) are on the last digit. Also shown are experimentally extrapolated values for the ground state energy \cite{Schnepp1970,Driessen1979}, as well as the value of the pressure at $T=4$ K, obtained by interpolating data in Table III of Ref. \cite{Driessen1979}.  }
  \label{t2}
  \begin{tabular*}{0.48\textwidth}{@{\extracolsep{\fill}}llll}
    \hline
    Potential &$e_K$ (K) & $E$ (K) & Pressure (bars) \\
     \hline
    Silvera-Goldman &$70.15(9)$ &$-87.94(4)$ &$-6.9(1)$ \\
    Buck &$71.37(5)$ &$-94.03(5)$ &$-26.6(2)$ \\
    Patkowski {\em et al.} &$78.52(5)$ & $-91.39(5)$ & $50.8(2)$\\
    Diep-Johnson &$78.44(7) $ &$-87.05(7)$ &$66.5(3)$ \\
        Experiment\cite{Schnepp1970} & &$-90.0$ &  \\
        Experiment\cite{Driessen1979} & &$-93.5$ & $\sim 6$\\
    \hline
  \end{tabular*}
\end{table}
At low temperature and at saturated vapor pressure, the equilibrium phase of {\em p}-H$_2$ is a hcp crystal, known experimentally \cite{fernandez2012}  to undergo virtually no thermal expansion  from $T=0$ all the way to the melting temperature, namely $T=13.8$ K, i.e., the density $n$ remains very close to its ground state value, namely \cite{Schnepp1970} $n=0.0261$ \AA$^{-3}$. Moreover, at $T=4$ K the system is very nearly in its ground state \cite{Omiyinka2013}. 
\\ \indent
In Table \ref{t2} we show the kinetic ($e_K$) and total ($E$) energy per molecule (in K), computed using various potentials; the results for the kinetic energy display the same trend already observed in the liquid phase, namely the {\em ab initio} potentials yield a value of the kinetic energy $\sim 10$\% higher than the SG and Buck potentials. Now, the total
energy per molecule for the SG and Buck potentials are in (nearly) perfect agreement with the $T=0$ theoretical estimate of Operetto and Pederiva \cite{Operetto2006}, who computed them by Diffusion Monte Carlo simulations. The value yielded by the P potential is $\sim$ 3.4 K lower than that of the SG, whereas that afforded by the Diep-Johnson potential is actually closer to that yielded by the SG potential (only a fraction of a K higher); on the other hand, the  Diep-Johnson and P potentials yield essentially identical estimates for $e_K$, i.e., the difference in the estimates of $E$ delivered by these two potentials is entirely due to the different attractive well depths. Comparison with experiment is again ambiguous, because in this case too, there exist two different estimates of the ground state energy, differing by roughly 3.5 K (see Table \ref{t2}). In any case, the estimate yielded by the P potential falls roughly between the two conflicting experimental values.
\\ \indent
Just like for the liquid phase, the SG potential yields a negative value of the pressure, while the P overestimates it by approximately 45 bars, based on an interpolation of results provided in Table III of Ref. \cite{Driessen1979}. The possible role of three-body terms in improving the agreement between theoretical and experimental results for the two potentials, both for the pressure and the energy,  is discussed in Sec. \ref{concl}.
\begin{figure}[h]
\centering
  \includegraphics[height=6cm]{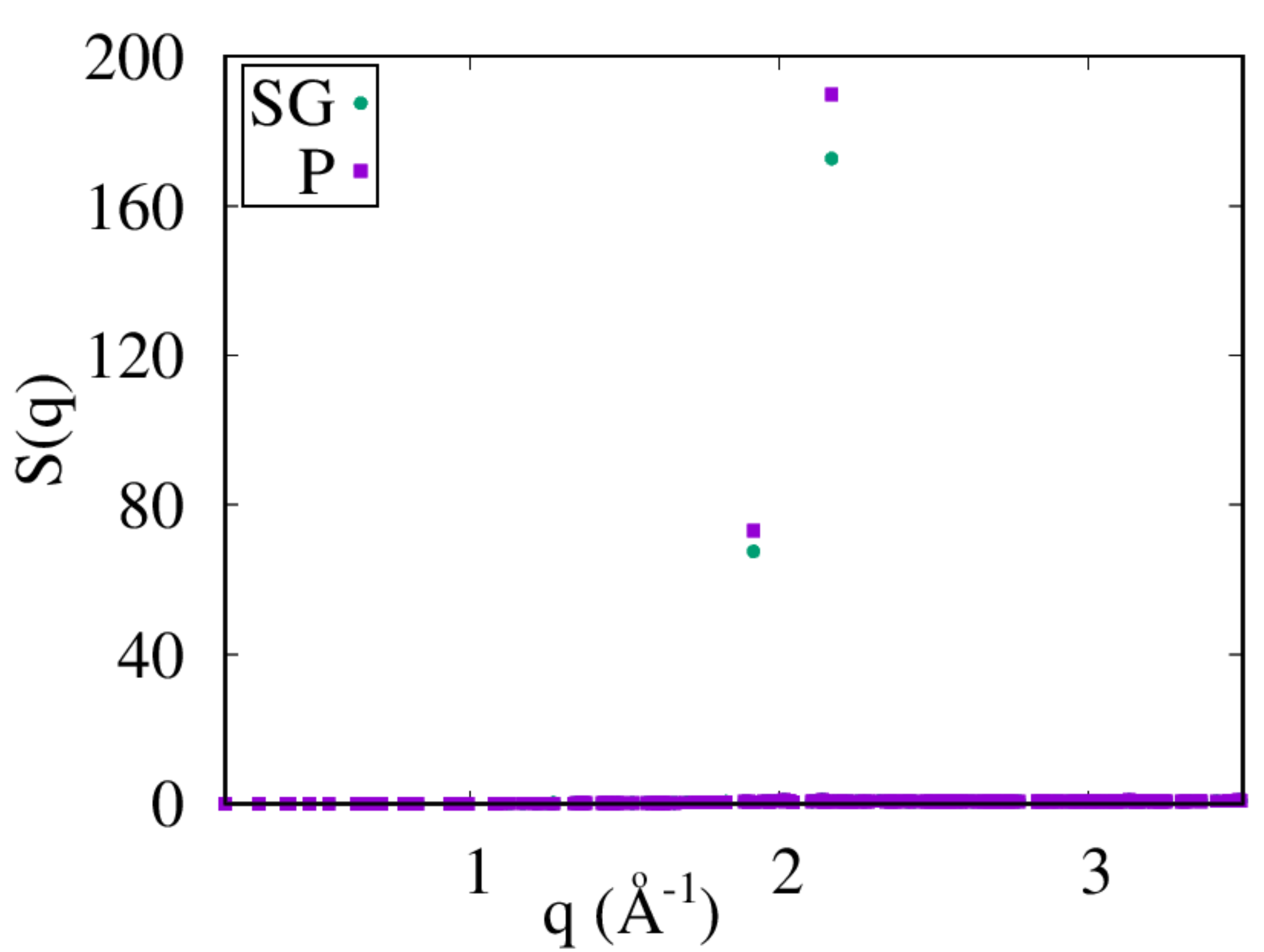}
  \caption{Static structure factor of hcp {\em p}-H$_2$ at density $n=0.0261$ \AA$^{-3}$ and temperature $T=4$ K, computed in this work using the SG (circles) and P (boxes) intermolecular pair potentials
  }
  \label{f2}
\end{figure}
\\ \indent 
Fig. \ref{f2} shows the static structure factor for the crystal, computed with the SG and P potentials, at the above-mentioned thermodynamic conditions. In this case, since the solid is not isotropic, $S(q)$ cannot be obtained from the pair correlation function but rather must be computed directly, for a selected set of wave vectors. Unlike in the liquid phase, as expected $S(q)$ displays in the crystal the characteristic, sharp (Bragg) peaks, whose strength greater for the P potential (by approximately 10\%), as observed in the liquid.
\section{Discussion and Conclusions}\label{concl} 
\indent
The Silvera-Goldman pair potential has been the most common choice for studies of the condensed phases of {\em p}-H$_2$, and in most studies of small {\em p}-H$_2$ clusters. In general, spherically symmetric pair potentials are utilized in physical situations in which one may not be necessarily pursuing chemical accuracy, but rather seek a reasonable numerical account of the most experimentally important thermodynamic properties of the condensed phases of molecular hydrogen. 
\\ \indent
In this work we have carried out first principle numerical simulations of liquid and solid {\em p}-H$_2$, based on the SG potential and on  other pair potentials, including some {\em ab initio} potentials developed over the past two decades. Our goal was that of carrying out an extensive comparison with experiment, in order to determine whether the SG remains the best choice of spherical pair potential. 
The results of our first principle numerical simulations show that the {\em ab initio} intermolecular pair potential by Patkowski {\em et al.} provides a  quantitatively more accurate description of the structure of the fluid, as shown by a comparison of computed and measured static structure factor.
\\ \indent
There are quantitative discrepancies between the estimates of the (kinetic) energy per molecule yielded by the SG and P potentials in the (liquid) solid phase. However, there are also significant, outstanding disagreements between different, independent experimental determinations of the kinetic energy in the liquid, as well as the energy in the crystal. It seems fair to state, however, that, in general, the {\em ab initio} potential of Patkowski {\em et al.} yields energy and pressure estimates closer to experiment than the Silvera-Goldman. The question arises of whether closer agreement with experiment can be achieved by including three-body terms.
\\ \indent
While the explicit inclusion of three-body contributions to the potential energy went beyond the scope of this work, some general considerations can be made, based on the experience accumulated with research on the condensed phases of helium. Three-body interaction potentials broadly include two distinct contributions, namely {\em a}) the (mainly repulsive) triple–dipole Axilrod–Teller (AT) term \cite{Axilrod1943}, and {\em b}) an attractive contribution,
arising from electronic exchange taking place in triplets of
neighboring molecules. The AT term has been shown to give a negligible contribution, at the thermodynamic conditions considered here\cite{Operetto2006}, and therefore one can expect the attractive part, effectively included in the SG potential, to play the most important role. 
\\ \indent
In the context of the condensed  phases of helium, three-body interactions have been shown \cite{Moroni2000,Chang2001} to have virtually no effect on the kinetic energy per particle, nor on structural correlations, which at least at moderate conditions of temperature and pressure are mostly sensitive to the (repulsive) short-range part of the pair potential. On the other hand, they can be expected to make the potential energy more negative (therefore possibly {\em worsening} the agreement with experiment yielded by the potential of Patkowski {\em et al.}) and to have a ``softening'' effect on the pressure, in this case improving the agreement with experiment \cite{Boninsegni1994}. Obviously, the situation may well be quantitatively different for molecular hydrogen, and progress can only be made by carrying out calculations explicitly including three-body terms, for which various forms have been proposed \cite{Wind1996,Hinde2008,Anatole2010,Huang2015,Ibrahim2022}.

\authorcontributions{The authors contributed equally to this work.}

\funding{This research was funded by the Natural Science and Engineering Research Council of Canada. Computing support from ComputeCanada is gratefully acknowledged.}

\dataavailability{The computer codes utilized to obtain the results can be obtained by contacting the authors.} 

\conflictsofinterest{The authors declare no conflict of interest.} 
\end{paracol}
\reftitle{References}
\externalbibliography{yes}
\bibliography{rsc}
\end{document}